# Transverse writing of three-dimensional tubular optical waveguides in glass with a slit-shaped femtosecond laser beam

Yang Liao[1], Jia Qi[1,2], Peng Wang[1,2], Wei Chu[1], Zhaohui Wang[1,3], Lingling Qiao[1], Ya Cheng[1,4*]

[1]State Key Laboratory of High Field Laser Physics, Shanghai Institute of Optics and Fine Mechanics, Chinese Academy of Sciences, Shanghai 201800, China

[2]School of Physical Science and Technology, Shanghai Tech University, Shanghai 200031, China

[3]University of Chinese Academy of Sciences, Beijing 100049, China

[4]Collaborative Innovation Center of Extreme Optics, Shanxi University, Taiyuan, Shanxi 030006, People's Republic of China

*ya.cheng@siom.ac.cn

# ABSTRACT

We report on fabrication of tubular optical waveguides buried in ZBLAN glass based on transverse femtosecond laser direct writing. Irradiation in ZBLAN with focused femtosecond laser pulses leads to decrease of refractive index in the modified region. Tubular optical waveguides of variable mode areas are fabricated by forming the four sides of the cladding with slit-shaped femtosecond laser pulses, ensuring single mode waveguiding with a mode field dimension as small as ~ 4 μm.

## Introduction

Optical waveguides that guide light along predetermined paths are a crucial building block of integrated photonic devices. Recently, fabricating optical waveguides with femtosecond laser direct writing has attracted much attention, as this approach uniquely allows formation of three-dimensional (3D) photonic circuits in various transparent materials (i.e., glass, crystals and polymers) in a continuous single-step processing manner [1-6]. In many glasses such as fused silica and borosilicate glass, waveguides written by focused femtosecond laser beams consist of cores with increased refractive index that overlap the center of focal spots [7]. Meanwhile, femtosecond laser can also induce negative refractive index change in several kinds of glass and crystals, such as ZBLAN glass, YAG crystal and lithium niobate crystal [8-12]. In these materials, waveguides are generally produced by forming a cladding region of reduced refractive index by irradiation of femtosecond laser pulses. The waveguides fabricated in such a way are usually termed "tubular waveguides" as the cladding appears like a tube which surrounds the waveguiding area unaffected by the laser irradiation. In the transverse writing scheme, the claddings can be formed by parallelly overlapping multiple tracks written by the focused femtosecond laser pulses along the boundaries of the waveguiding areas [8,9]. This requires a considerable number of scan times to form the claddings, and the large thicknesses of the formed claddings may cause considerable difficulty in using such waveguides for constructing evanescent-coupling-based photonic circuits such as beam splitters and Mach-Zender interferometers. Alternatively, the tubular waveguides can also be formed with a longitudinal writing scheme [13-15]. In principle, this approach allows formation of tubular waveguides with small mode areas and thin cladding layers provided that a tight focusing condition is chosen by use of high-numerical-aperture (high-NA) focal lenses. However, the longitudinal writing scheme inherently suffers from lack of flexibility in terms of writing 3D photonic circuits due to the limited working distances of high-NA lenses.

In this article, we demonstrate transversely writing the tubular waveguides by focusing slit-shaped femtosecond laser beams in glass to form the tubular claddings. Single-mode propagation in the fabricated waveguides has been achieved, providing variable mode field dimensions as small as ~ 4 μm.

**Results and Discussion**

In this work, $ZrF_4$-$BaF_2$-$LaF_3$-$AlF_3$-$NaF$ (ZBLAN) fluoride glass was used as the substrate. It has been reported by Lancaster et al. that the refractive index change induced by femtosecond laser irradiation in ZBLAN glass can reach a value between ~ -1.5×$10^{-3}$ to ~ -1×$10^{-3}$, which is sufficient to form the cladding for constructing tubular waveguides [9]. The experimental setup is illustrated in Fig. 1. The experimental details are provided in **Methods**. Here, to facilitate the flexible spatial beam shaping, we used the same layout as reported by Salter et al. [16].

To optimize the focusing parameters, numerical simulations were performed. Figure 2 presents the simulated light distributions of a femtosecond laser pulse focused in glass after first passing through a slit as schematically illustrated in Fig. 1 and then an air objective of NA=0.80. Considering that the refractive index of the ZBLAN glass is ~1.5, an NA of 0.80/1.5=0.53 was chosen in our simulation. The simulations of the intensity distributions of the focal spots in the Y-Z plane (see, the coordinates indicated in Fig. 1) were performed using Fresnel diffraction formula [17],

$$I(x,y,z) = \left| \frac{exp\,(ikz)}{i\lambda z} \iint_{-\infty}^{\infty} E_1\,(\xi,\eta) exp\,(-ik\frac{\xi^2+\eta^2}{2f}) exp\,(ik\frac{(x-\xi)^2+(y-\eta)^2}{2z}) d\xi d\eta \right|^2 \quad (1)$$

where $E_1$ is the light field at the entrance aperture of the objective lens, $k$ and $\lambda$ are the wave number and wavelength of the light in the glass, and $f$ is the focal length of the objective lens. It should be noted that to use the above formula, we have replaced the slit-shaped beam with an elliptical Gaussian beam whose waists along the long and short axes are the same as the length and width of the slit, which can be expressed as

$$E_1(x,y) = E_0 exp[-\left(\frac{x^2}{S_x^2} + \frac{y^2}{S_y^2}\right)] \quad (2)$$

where $E_0$ is the constant field amplitude, $2S_x$ and $2S_y$ are the slit size along the x axis and y axis, respectively. Such an approximation significantly simplifies the calculation whilst provides the similar results as that obtained with rigorous diffraction theory [18,19].

The results in Fig. 2 show that by properly choosing the dimensions of the slit, the slit beam shaping method can produce highly elliptical focal spots in the cross-sectional plane with either very high or very low aspect ratios. Thus, one can arrange such focal spots to construct the claddings of tubular waveguides. The horizontal sides of tubular claddings shown in Figs. 2(a) and 2(d) are generated with a 200-μm-width slit and a 160-μm-width slit, respectively, and the vertical sides shown in Figs. 2(b) and 2(e) are generated with a 2000-μm-width slit and a 2400-μm-width slit, respectively. As illustrated in Fig. 2(c), a tubular waveguide with a cross section size of 6 μm × 6 μm can be produced by overlapping the structures induced by the focal spots in Figs. 2(a) and (b); whereas the tubular waveguide with a cross section size of 12 μm × 12 μm as illustrated in Fig. 2(f) can be formed with the focal spots in Figs. 2(d) and (e). Thus, tubular waveguides can be buried in various transparent materials using this scheme with a large tuning range of size and aspect ratio of the cross section. We would like to stress that the above simulations are carried out mainly as a conceptual demonstration. Meanwhile, the results in Fig. 2 also provide guidance in optimizing the experimental parameters. One must realize that many effects such as the nonlinear multiphoton ionization, self-focusing, as well as optical aberration are not taken into account in the above simulations [20-23]. For this reason, the optimum experimental parameters of the slits used below are obtained by trial and error process, which are not exactly the same as that chosen above in the simulations.

Figure 3(a) shows a tubular waveguide of a cross sectional size of 6 × 6 $μm^2$ written 100 μm beneath the glass surface. To fabricate the two vertical sides of the square-shaped cladding, a laser pulse energy of 1 μJ was chosen, and the width of the slit on the SLM were set to 2000 μm, respectively. In contrast, the two horizontal sides of the cladding were written with 4.9 μJ pulses by setting the width of the slit to 200 μm, respectively. The scan speed for writing both the vertical and horizontal sides of the cladding was 50 μm/s. The near-field intensity distributions captured by CCD camera are presented in Figs. 3(b) and 3(c), showing clearly a single mode profile. To estimate the propagation loss in the waveguide, we captured the top-view image of the waveguide under a microscope when it was carrying the propagating He-Ne laser beam, as shown in Fig. 4(a). From the exponential decay of the scattering light along the waveguide, the loss in the waveguide is estimated to be ~3.3 dB/cm. At first glance, this loss appears high as the typical loss of waveguides written in fused silica is only ~0.5 dB/cm [3]. However, we would like to point out that there is a great potential to minimize the loss by optimizing the fabrication conditions in the future. As have been discussed by Lancaster et al. in Ref. [9], the loss in the tubular waveguides written in ZBLAN glass by femtosecond laser is synergistically determined by the thickness and the refractive index change of the cladding as well as the size of mode area. Considering the fact that the thickness of the two vertical sides of the square-shaped cladding is only ~ 4 μm, we found that our measured loss qualitatively agrees well with the modeling results by Lancaster et al. [see, Fig. 3 in Ref. 9]. Indeed, as we show below, the loss is significantly reduced in a tubular waveguide of a larger mode area. In the future, we are going to reduce the loss by optimizing the thickness of the cladding in a systematical way. Meanwhile, inducing higher refractive index changes with femtosecond laser irradiation may be achievable in other glass or crystalline materials, which also enables low-loss propagation in the tubular waveguides.

To experimentally demonstrate the wide range of tuning flexibility, we fabricated another tubular waveguide of a larger cross sectional size $12 \times 12$ μm$^2$ buried 100 μm beneath the glass surface, as shown in Fig. 3(d). The lengths of the four sides of the cladding can be controlled by varying either the NA of objective lens or the dimensions of the slits. We took the latter option because this requires no mechanical realignment of the optical system. Furthermore, the size of the laser affected zone is also influenced by the laser pulse energy due to the nonlinear self-focusing effect, which provides another control knob for optimizing the dimensions of the claddings. To write the two horizontal sides of the cladding, the width of the slit on the SLM were changed to 160 μm, respectively. The pulse energy was changed to 5.6 μJ to ensure a sufficiently high intensity at the focus comparable to that in Fig. 3(a). Meanwhile, the two vertical sides of the cladding were written by setting the width of the slit to 2400 μm, respectively. Accordingly, the pulse energy was changed to 1.6 μJ. The scan speed for writing both the vertical and horizontal sides of the cladding was set at 50 μm/s. Unlike the single-mode profile as presented in Fig. 3(c), we observed a high-order mode along the vertical direction in the waveguide of larger cross sectional size as shown in Fig. 3(e) and 3(f). This indicates that the refractive index in the two horizontal sides of the cladding is too high to forbidden the high-order-mode propagation but only allow the single-mode propagation in the fabricated waveguide. In practice, one can achieve single-mode tubular waveguides at increased mode field sizes by reducing the refractive index change induced by the femtosecond laser irradiation, which can be easily realized by varying the deposited irradiation dose (i.e., laser pulse energy and/or scan speed). The loss estimated from measuring the decay in the scattering light from the top-view image of the waveguide is ~1.0 dB/cm, which is much less than that measured in the tubular waveguide in Fig. 3(a) and is consistent with the modeling result in Ref. [9].

At last, we performed quantitative analysis of the mode profile in Fig. 3(c), as shown in Fig. 4(b). The transverse and vertical profiles of the mode reveals a mode field size of $3.9 \times 4.2$ μm$^2$ (FWHM). Although in our design, the four sides of the cladding have a same length, which should in principle ensure a symmetric mode profile, the differences in the refractive indices and the thicknesses of the horizontal and vertical sides lead to the different capabilities to confine the light along these directions, giving rise to a slightly asymmetric mode profile. The issue can be overcome in a straightforward manner as our technique based on the laser direct writing provides a high flexibility in controlling the cross sectional size of the waveguide through changing of the distances between the two parallel sides in the vertical and horizontal directions.

## Conclusion

To conclude, we have demonstrated fabrication of tubular waveguides buried in glass using transverse femtosecond laser writing scheme. Based on the slit-beam shaping technique developed by Cheng et al. [18], we show that tubular waveguides of variable cross sectional dimensions can be achieved to realize either single or multi-mode propagation. Our technique is beneficial for constructing 3D photonic circuits in various transparent materials whose refractive indices are subject to decrease with the femtosecond laser irradiation. The tubular waveguides inherently prevent the waveguiding areas from modification by the laser irradiation, thereby maintaining all the linear and nonlinear optical properties of the pristine materials. Full exploitation of the unique advantage may lead to waveguide-based 3D photonic circuits of extremely low propagation loss and large nonlinear optical coefficients.

## Methods

**Fabrication of tubular optical waveguides.**

Home-made ZrF$_4$-BaF$_2$-LaF$_3$-AlF$_3$-NaF (ZBLAN) fluoride glass of a size of 10 × 10 × 3 mm$^3$ and polished on all six sides were used as the substrates. Figure 1 shows the schematic of our experimental setup. The output beam of a Ti:Sapphire laser (Libra-HE, Coherent Inc.) with a maximum pulse energy of 3.5 mJ, an operation wavelength of 800 nm, a pulse width of ~50 fs, and a repetition rate of 1 kHz was used. The pulse energy was controlled using a rotatable half-wave plate and a Glan-Taylor polarizer. The laser beam was first expanded using a concave lens of focal length 10 cm (L1) and a convex lens of focal length 20 cm (L2) before impinging on a reflective phase-only spatial light modulator (SLM, Hamamatsu, X10468-02). To form the slit-shaped beam with the phase-only SLM, a blazed grating with a modulation depth of 2π rad was used to maximize the diffraction efficiency of the first order. Example phase masks used for writing the horizontal and vertical sides of the cladding are presented in Figs. 1(b) and (c), respectively. The first diffracted order produced by the SLM, after being focused with a lens of focal length 70 cm (L3), was filtered out using a pinhole. To obtain a sufficiently large diffraction angle, the period of the gratings in Figs. 1(b) and (c) were set to 420 μm and 100 μm, respectively. The slit-shaped beam were then imaged onto the back aperture of the objective lens (MPLFLN, Olympus) using a lens of focal length 70 cm (L4), and focused into the glass sample to write the buried waveguides. The glass sample was mounted on a computer-controlled XYZ stage with a translation resolution of 1 μm. To fabricate the enclosed claddings, single scans of slit-shaped beams at a fixed scan speed of 0.02 mm/s were performed to write the four sides of the square-shaped claddings in such an order that the earlier fabricated structures will not influence the propagation of the laser beam in the latter scans. The fabrication parameters for writing the waveguides of different dimensions are described below in detail when the fabricated structures will be introduced.

**Characterization of Guidance.**

The guiding properties of the fabricated tubular waveguides are explored under a typical end-face coupling arrangement. The linear polarized 632.8 nm He-Ne laser is used as the light source. It is focused by a 10 × microscope object lens (NA = 0.30) and then coupled into the entrance of the waveguide. Another 20 × microscope object lens (NA = 0.40) is utilized as the out-coupler, through which the modal profile at the output of the structure is collected and recorded by a CCD camera.

To estimate the propagation loss in the waveguide, the exponential decay of the scattered light along the optical waveguide carrying the He-Ne laser was recorded with a top-view CCD camera. After recording the scattered intensity along the waveguide, an exponential fit to the data was used to determine the propagation loss. To avoid the influence of strong scattering at the input and output facets on measurement accuracy, a 5-mm-long streak scattered out of the middle area of the waveguide was selected for loss evaluation.

# References

1. Davis, K. M., Miura, K., Sugimoto, N. Hirao, K. Writing waveguides in glass with a femtosecond laser. *Opt. Lett.* **21**, 1729-1731 (1996).

2. Itoh, K., Watanabe, W., Nolte, S. & Schaffer, C. B. Ultrafast processes for bulk modification of transparent materials. *MRS Bulletin* **31**, 620-625 (2006).

3. Ams, M. *et al.* Ultrafast laser written active devices. *Laser & Photon. Rev.* **3**, 535-544 (2009).

4. Osellame, R., Hoekstra, H. J., Cerullo, G. & Pollnau, M. Femtosecond laser microstructuring: an enabling tool for optofluidic lab-on-chips. *Laser Photonics Rev.* **5**, 442-463 (2011).

5. Sugioka, K. & Cheng, Y. Ultrafast lasers—reliable tools for advanced materials processing. *Light: Sci. Appl.* **3**, e149 (2014).

6. Eaton, S. M. *et al.* Transition from thermal diffusion to heat accumulation in high repetition rate femtosecond laser writing of buried optical waveguides. *Opt. Express* **16**, 9443-9458 (2008).

7. Miura, K. *et al.* Photowritten optical waveguides in various glasses with ultrashort pulse laser. *Appl. Phys. Lett.* **71**, 3329-3331 (1997).

8. Okhrimchuk, A. G., Shestakov, A. V., Khrushchev, I. & Mitchell, J. Depressed cladding, buried waveguide laser formed in a YAG:$Nd^{3+}$ crystal by femtosecond laser writing. *Opt. Lett.* **30**, 2248-2250 (2005).

9. Lancaster, D. G. *et al.* Fifty percent internal slope efficiency femtosecond direct-written $Tm^{3+}$:ZBLAN waveguide laser. *Opt. Lett.* **36**, 1587-1589 (2011).

10. Thomson, R. R. *et al.* Optical waveguide fabrication in z-cut lithium niobate using femtosecond pulses in the low repetition rate regime. *Appl. Phys. Lett.* **88**, 111109 (2006).

11. Liao, Y. *et al.* Electro-optic integration of embedded electrodes and waveguides in LiNbO$_3$ using a femtosecond laser. *Opt. Lett.* **33**, 2281-2283 (2008).

12. Chen F. & Vazquez de Aldana, J. R. Optical waveguides in crystalline dielectric materials produced by femtosecond-laser micromachining. *Laser Photonics Rev.* **8**, 251-275 (2014).

13. Long, X. *et al.* Stressed waveguides with tubular depressed-cladding inscribed in phosphate glasses by femtosecond hollow laser beams. *Opt. Lett.* **37**, 3138-3140 (2012).

14. Beckmann, D. *et al.* Beam shaping of laser diode radiation by waveguides with arbitrary cladding geometry written with fs-laser radiation. *Opt. Express* **19**, 25418-25425 (2011).

15. Caulier, O., Le Coq, D., Bychkov, E. & Masselin, P. Direct laser writing of buried waveguide in As$_2$S$_3$ glass using a helical sample translation. *Opt. Lett.* **38**, 4212-4215 (2013).

16. Salter, P. S. *et al.* Adaptive slit beam shaping for direct laser written waveguides. *Opt. Lett.* **37**, 470-472 (2012).

17. Goodman, J. Introduction to Fourier Optics (Roberts & Company, Englewood, Colorado, 2005).

18. Cheng, Y. *et al.* Control of the cross-sectional shape of a hollow microchannel embedded in photostructurable glass by use of a femtosecond laser. *Opt. Lett.* **28**, 55-57 (2003).

19. Ams, M., Marshall, G. D., Spence, D. J. & Withford, M. J. Slit beam shaping method for femtosecond laser direct-write fabrication of symmetric waveguides in bulk glasses. *Opt. Express* **13**, 5676-5681 (2005).

20. Bloembergen, N. A brief history of light breakdown. *J. Nonlinear Opt. Phys.* **6**, 377-385

(1997).

21. Couairon A. & Mysyrowicz, A. Femtosecond filamentation in transparent media. *Phys. Rep.* **441**, 47-189 (2007).

22. Mauclair, C. *et al.* Ultrafast laser writing of homogeneous longitudinal waveguides in glasses using dynamic wavefront correction. *Opt. Express* **16**, 5481–5492 (2008).

23. Jesacher, A., Marshall, G. D., Wilson, T. & Booth, M. J. Adaptive optics for direct laser writing with plasma emission aberration sensing. *Opt. Express* **18**, 656–661 (2010).

## Acknowledgments

We gratefully acknowledge Dr. Liyan Zhang and Prof. Haiwen Cai for their contributions to this research. This work was supported by the National Basic Research Program of China (Grant No. 2014CB921300), National Natural Science Foundation of China (Grant Nos. 61590934, 11134010, and 61327902), and the Youth Innovation Promotion Association of Chinese Academy of Sciences.

## Author contributions

Y. C. and Y. L. planned and designed the experiments. Y. L., J. Q., and P. W. performed the experiment. Z. W. and J. Q. performed the simulation. Y. C. and Y. L. wrote the paper. All authors participated in the discussion of the results.

## Additional information

Competing financial interests: The authors declare no competing financial interests.

**Figure Captions:**

Figure 1 Schematic illustration of the experimental setup. POL: polarizer. CCD: charge coupled device. OBJ: objective lens. PC: personal computer. L1, L2, L3, and L4 are the lenses of different focal lengths which are described in the main text. Coordinates are indicated in the figure. Inset: An example of phase mask for writing the horizontal sides of the cladding.

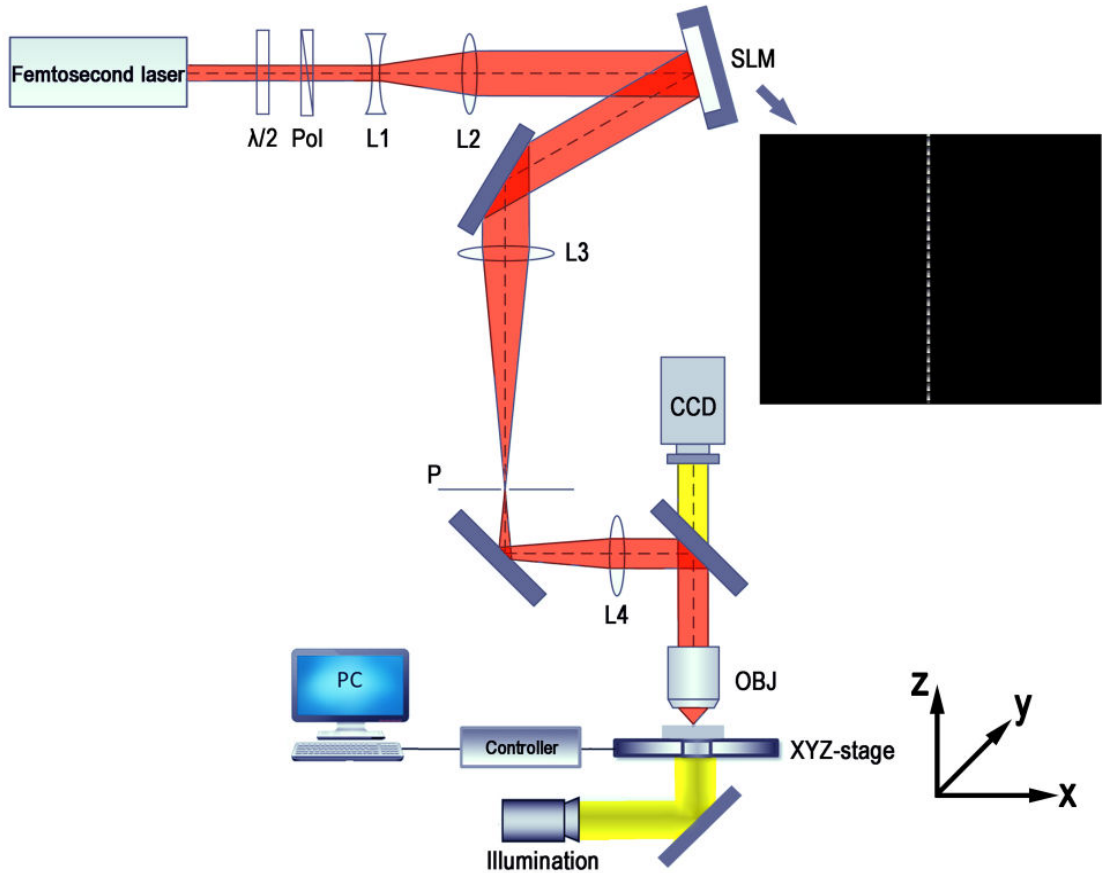

Figure 2 Simulated distributions of peak laser intensity in the focal spots of slit-shaped beams. For fabrication of small-mode-area waveguide, the widths of the slits in (a) and (b) are 200 μm, and 2000 μm, respectively. For fabrication of large-mode-area waveguide, the widths of the slits in (d) and (e) are 160 μm, and 2400 μm, respectively. (c) A 12 μm × 12 μm tubular waveguide can be produced by overlapping the structures shown in Figs. 2(a) and (b); (f) A 6 μm × 6 μm tubular waveguide can be produced by overlapping the structures shown in Figs. 2(d) and (e).

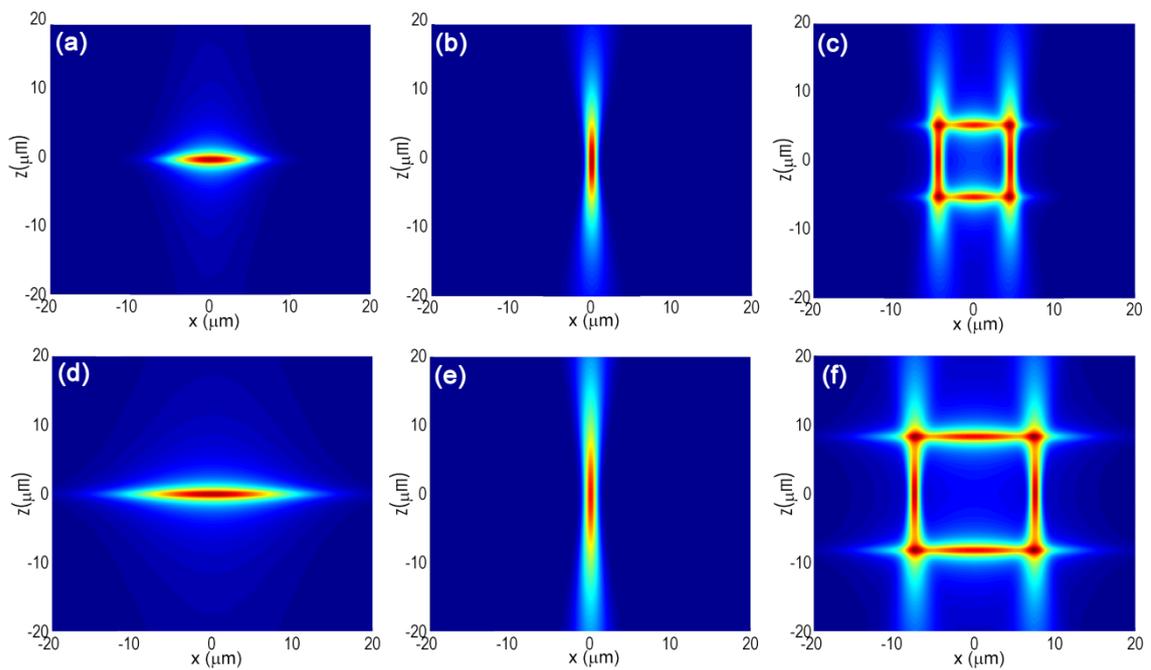

Figure 3 (a) Optical micrograph of a fabricated tubular waveguides with a cross sectional size of 6 × 6 μm$^2$; (b) Near-field intensity distribution under bright field showing the position of guiding mode in the single mode waveguide; (c) Near-field intensity distribution under dark field measured at the exit of the single mode waveguide; (d) Optical micrograph of a tubular waveguides with a cross sectional size of 12 × 12 μm$^2$; (e) Near-field intensity distribution under bright field showing the position of guiding mode in the multimode waveguide; (f) Near-field intensity distribution under dark field at the exit of the multimode waveguide; Scale bar: 10 μm.

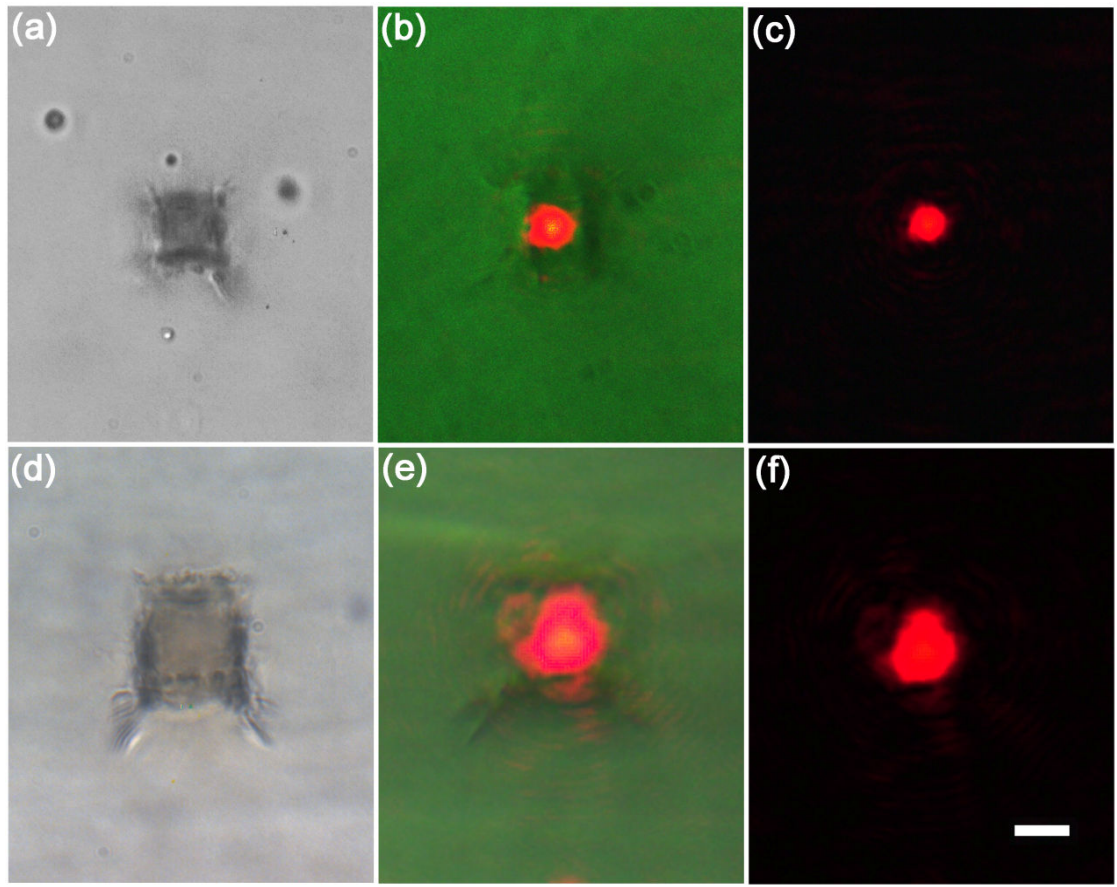

Figure 4 (a) Top view images of a 10-mm-long waveguide showing the decay of scattered light along the waveguide. (b) 2D plot of the intensity distribution of the mode in the waveguide with a cross sectional size of 6 × 6 μm$^2$. The vertical and transverse profiles of the mode are presented, respectively.

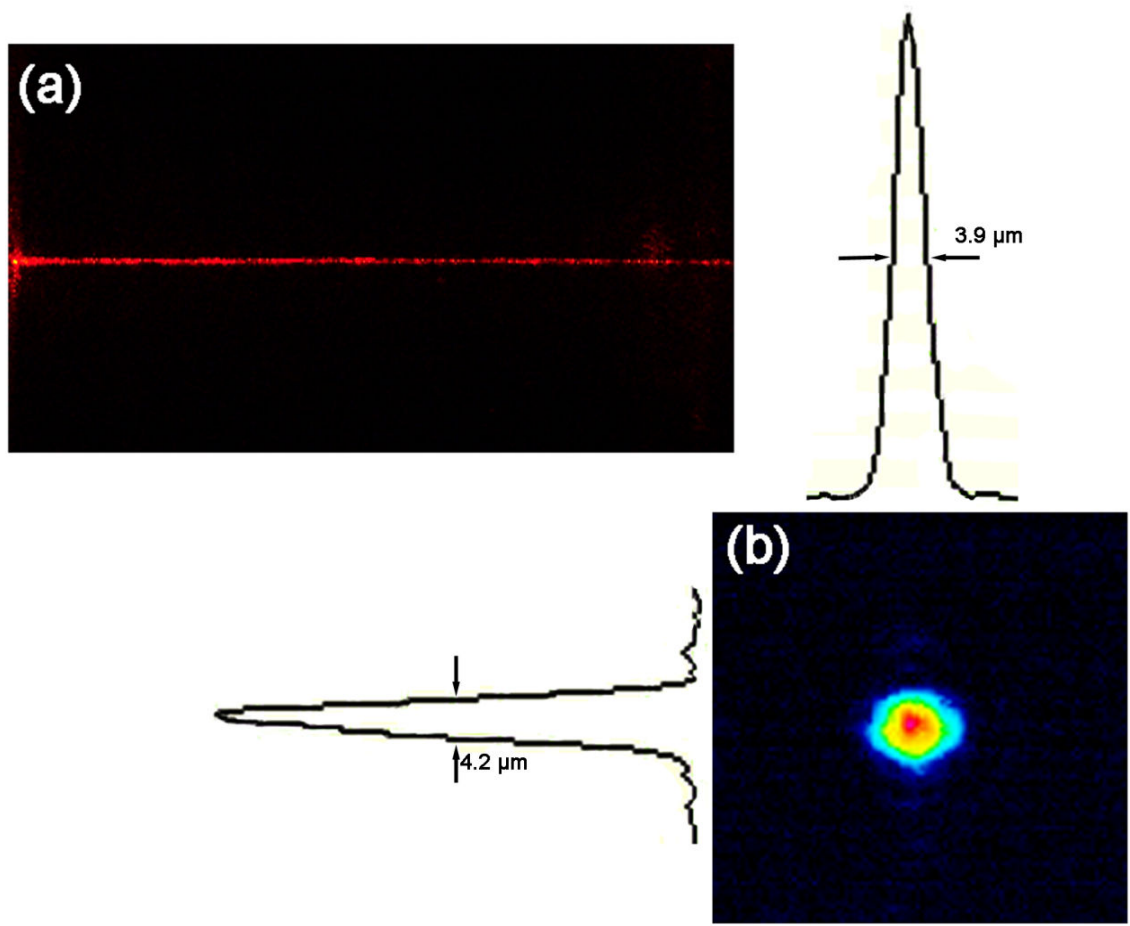